# Coexistence of multiple silicene phases in silicon grown on Ag(111)

P. Moras[1], T. O. Mentes[2], P. M. Sheverdyaeva[1], A. Locatelli[2], and C. Carbone[1]

[1]Istituto di Struttura della Materia, Consiglio Nazionale delle Ricerche, Trieste, Italy
[2]Sincrotrone Trieste S.C.p.A, SS 14, Km 163,5, I-34149, Trieste, Italy

Silicene, the silicon equivalent of graphene, is attracting increasing scientific and technological attention in view of the exploitation of its exotic electronic properties. This novel material has been theoretically predicted to exist as a free-standing layer in a low-buckled, stable form, and can be synthesized by the deposition of Si on appropriate crystalline substrates. By employing low-energy electron diffraction and microscopy, we have studied the growth of Si on Ag(111) and observed a rich variety of rotationally non-equivalent silicene structures. Our results highlight a very complex formation diagram, reflecting the coexistence of different and nearly degenerate silicene phases, whose relative abundance can be controlled by varying the Si coverage and growth temperature. At variance with other studies, we find that the formation of single-phase silicene monolayers cannot be achieved on Ag(111).

# 1. Introduction

Silicene is a single layer of Si atoms arranged in a honeycomb lattice, expected to be stable in a low-buckled, free-standing geometry [1]. Theoretical calculations for the silicene electronic structure predict that the π and π* bands cross linearly the Fermi level near the K and K' points of the surface Brillouin zone, the charge carriers behaving as massless Dirac fermions [1-4]. These studies indicate that free-standing silicene presents several analogies to graphene, with differences mainly arising from the larger atomic size and spin-orbit of silicon compared to carbon. Due to the spin-orbit coupling, the quantum spin Hall effect, predicted for graphene [5], should be sizably stronger in silicene [6,7]. Also, field-gated silicene could enable electrically switched spin-filter [8].

Although silicene integration in Si-based device architectures is attractive, a viable path for the production of silicene layers with uniform structural properties and unperturbed electronic states is still to be established. Recently, several experimental studies have addressed the epitaxial synthesis of silicene on various supporting substrates. Except for the reports on silicene formation on $ZrB_2$/Si(111) [9,10] and Ir(111) [11], most studies focused on silicene synthesis on Ag(111) [12-22]. Scanning tunnelling microscopy (STM) and low-energy electron diffraction (LEED) investigations identify distinct silicene structures on Ag(111), with characteristic reconstructions and significant out of plane buckling (0.5-0.8 Å) [13-19]. First principles calculations indicate, that Si on Ag(111) can give origin to nearly degenerate distinct silicene phases with distorted honeycomb atomic arrangement and different rotational alignment with respect to the substrate [23]. These Si structures appear to reflect the existence of various types of atomic ordering, with partial $sp^2$ and $sp^3$ configurations, that are stabilized on Ag(111). Figure 1 presents simple structural models (top row), and corresponding LEED patterns (bottom row), for different silicene phases on Ag(111), with (4×4), ($\sqrt{13}\times\sqrt{13}$)R13.9° and ($2\sqrt{3}\times2\sqrt{3}$)R30° reconstructions [12-15,17,19]. They are characterized by different in-plane rotations of the silicene lattice with respect to the Ag[110] axis, which leads to slight structural changes determined by the lattice matching conditions to the Ag(111) surface [13]. The notation ($n\times n$)Rβ is used to identify size and orientation of the silicene supercells, with respect to the Ag(111) surface unit cell and the Ag[110] axis. All these silicene structures, except the (4×4), form two equivalent mirror domains with respect to the Ag[110] direction.

The conditions for the growth of different silicene structures and their description significantly vary in the literature. The formation of a complete single-phase monolayer (ML) is reported for the (4×4) [13,14,17,19] and ($2\sqrt{3}\times2\sqrt{3}$)R30° structures [12,13]. The single-phase (4×4) silicene is found to form at 250-290 °C [19], ~ 220-260 °C [17], and 150 °C [13]. It is also shown that at the monolayer completion the (4×4), ($\sqrt{13}\times\sqrt{13}$)R13.9° and ($2\sqrt{3}\times2\sqrt{3}$)R30° domains coexist under certain growth conditions [13,14,19]. The lattice constant for the ($2\sqrt{3}\times2\sqrt{3}$)R30° monolayer structure, determined by STM [12], is questioned in Ref. [24]. Some STM and theoretical calculations attribute the ($\sqrt{13}\times\sqrt{13}$)R13.9°$a$ phase to a very strongly compressed silicene, with density higher by ~ 20% than the (4×4) structure [14,19]. At high deposition temperature a different monolayer silicene structure is claimed to form, consisting of equivalent but azimuthally rotated domains of ($\sqrt{3}\times\sqrt{3}$)R30° periodicity, with reference to an ideal and unperturbed silicene lattice [16,18,20]. This contradicts other works that assign the ($\sqrt{3}\times\sqrt{3}$)R30° structure to a coverage in excess of 1 ML [19,25]. In consideration of the contrasting reports, it appears that several aspects of the silicene formation as well as the conditions for the synthesis of a specific phase call to be better established.

In this paper we examine the growth of silicene on Ag(111), with focus on the formation sequence of different phases and the conditions for their coexistence. We determine the relative phase abundance depending on temperature and coverage. We address in particular the issue whether a single-phase monolayer exists, and show that, under the examined growth conditions, the formation of a single-phase silicene monolayer does not take place. Our results reveal instead a complex, kinetically limited growth scenario, where the simultaneous formation of different silicene phases competes with multilayer Si growth and surface dewetting processes.

## 2. Experimental

The experiments were performed using the SPELEEM instrument at the Nanospectroscopy beamline of the Sincrotrone Trieste, Elettra [26]. The microscope was used in either microprobe-LEED or low-energy electron microscopy (LEEM) mode for real-time monitoring of Si growth on Ag(111). Microprobe diffraction operation was obtained by imaging the back-focal plane of the objective lens, after inserting appropriate apertures restricting the illumination beam to a minimum diameter of 0.5 microns. Direct real-space imaging with LEEM allowed a lateral resolution of about 10 nm. The field-of-view, ranging from 2 μm to 30 μm, gave the possibility to probe morphology from mesoscopic to microscopic length scales. Bright field or dark field images were obtained from the specular (00) beam or higher diffraction orders, respectively, selected by placing a circular aperture at the diffraction plane [27]. The base pressure of the microscope chamber was $1\times10^{-10}$ mbar.

The Ag(111) substrate was treated with standard sputtering-annealing cycles to produce a sharp (1×1) LEED pattern. LEEM images of the surface revealed a distribution of atomically flat terraces (100 nm average width) separated by one atom high steps. Step bunches with a much stronger contrast than single atom steps were only occasionally found. Si was deposited by resistive annealing of a Si wafer at rates from 0.01 to 0.05 ML/min, in order to have the same rate used in previous works [13,14,18,19,22]. During the Si evaporation the pressure increased to $4\times10^{-10}$ mbar. 1 ML Si was determined as the quantity corresponding to the completion of the Si wetting layer. This definition was cross checked with other real and reciprocal space features, as discussed in the following. In-situ core level photoemission spectroscopy (not shown) was used to monitor the level of substrate and silicene contamination, which remained below the detection limit (less than 0.1% of a ML) during the measurements. The calibration of the W-Re thermocouple in the sample holder was performed in high vacuum conditions (~$1\times10^{-7}$ mbar) through an ancillary thermocouple, that was contacted with the Ag crystal surface and removed before the bake-out procedure of the microscope chamber.

## 3. Results and Discussion

Several Si growth sequences at different and well controlled substrate temperatures (determined with an experimental uncertainty of ± 10 °C) were carried out in order to map the complex phase formation diagram of the system. The temperature for the Si deposition was varied in the 135-315 °C range to reproduce the growth conditions reported in earlier works. Figure 2 presents the LEED and LEEM data obtained after deposition of 1.3 ML Si on Ag(111) at 248 °C. The LEED pattern in Fig. 2(a) originates from a 6 μm wide surface region and shows that the (4×4) and ($\sqrt{13}\times\sqrt{13}$)R13.9°$a$ phases coexist at this temperature and coverage. Dark field microscopy (Fig. 2(b-d)) displays the monolayer domain structure, by resolving the lateral distribution of the (4×4) and ($\sqrt{13}\times\sqrt{13}$)R13.9°$a$ structures. The dark field images are obtained by selectively filtering the first order diffraction spots of the respective silicene phases (Fig. 1(e,f)). The domains of the ($\sqrt{13}\times\sqrt{13}$)R13.9°$a$ phase (Fig. 2(c,d)) have a mean size of 215 ± 14 nm and cover 60 ± 3% of the substrate. The domains of the (4×4) phase (Fig. 2(b)) have a mean size of about 200 nm with a large variation and cover 23 ± 5% of the Ag surface. The sum of the dark field images Fig. 2(b-d), reported in Fig. 2(e), covers nearly the entire surface, leaving small dark gray regions, that are found to correspond closely to the distribution of white patches in the bright field LEEM image of Fig. 2(f). This image illustrates the morphology of the silicene monolayer (light gray), covering the almost entire Ag(111) substrate. Additional small dark gray (15% of the surface) and larger white areas (5% of the surface) are associated with a ($\sqrt{3}\times\sqrt{3}$)R30° diffraction pattern (a fractional spot is indicated by an arrow in Fig. 2(a)). This observation indicates that Si in excess of 1 ML gives origin to ($\sqrt{3}\times\sqrt{3}$)R30° terminated islands, in accordance with Refs. [19,25]. Indeed, as we will report in the following, the appearance of the ($\sqrt{3}\times\sqrt{3}$)R30° pattern coincides with the closure of the silicene wetting layer, and marks the monolayer coverage. The coverage calibration thus obtained has an experimental uncertainty as low as ±5%.

Importantly, the principal observation that the Si growth produces a structurally heterogeneous surface turns out to be the case for all investigated temperatures in the sub-monolayer and monolayer regimes, with varying relative amounts of different silicene phases depending on the temperature and coverage. This aspect is illustrated in Figure 3, which displays the LEED patterns for 1 ML Si grown at different temperatures. All diffraction patterns originate from mixtures of two or more different silicene phases. As a general trend, increasing temperature results in sharper diffraction spots, thus indicating larger domains for each phase, and less rotational broadening. The entire temperature range at the monolayer coverage can be divided into three regimes: i) up to 248 °C the (4×4), ($\sqrt{13}\times\sqrt{13}$)R13.9°$a$ and $b$ phases are always coexisting (see Fig.3(a-d)) [28]; ii) at about 268 °C (see Fig. 3(e)), within a temperature window of less than 40 °C, the (4×4), ($\sqrt{13}\times\sqrt{13}$)R13.9°$a$ and ($2\sqrt{3}\times2\sqrt{3}$)R30° structures coexist [29], whereas the ($\sqrt{13}\times\sqrt{13}$)R13.9°$b$ spots are very weak; iii) above 290 °C (Fig. 3(f)) the pattern is dominated by the ($2\sqrt{3}\times2\sqrt{3}$)R30° spots, with minor contribution from the (4×4) phase.

The coexistence of different silicene phases characterizes all stages of the Si growth for the entire temperature range we explored. Figure 4 shows the coverage dependence of the LEED patterns for the growth sequences performed at 218 °C (bottom row), 268 °C (central row) and 296 °C (top row) up to 1.3 ML Si. The right column of Fig. 4 summarizes the intensity evolution of the LEED spots for the different silicene phases. First, the ($\sqrt{3}\times\sqrt{3}$)R30° spot intensity (measured on fractional spots similar to that indicated in Fig. 2(a)) appears only above 1 ML for all temperatures. This observation excludes that the ($\sqrt{3}\times\sqrt{3}$)R30° pattern can be associated with a monolayer structure in direct contact with Ag. The ($\sqrt{13}\times\sqrt{13}$)R13.9°$b$ phase, with relatively broad diffraction spots, is present only in the low temperature regime. Here the (4×4) and ($\sqrt{13}\times\sqrt{13}$)R13.9°$a$ structures have similar coverage dependence. At higher temperatures, the (4×4) phase appears somewhat earlier in comparison to the other competing structures, namely ($\sqrt{13}\times\sqrt{13}$)R13.9°$a$ and ($2\sqrt{3}\times2\sqrt{3}$)R30°, but the intensity of the latter two phases becomes dominant with increasing coverage. This is most evident at the highest temperature, where the (4×4) phase nucleates at lower coverage, but saturates as soon as the ($2\sqrt{3}\times2\sqrt{3}$)R30° structure starts growing. In addition to the Si phases, the (1×1) Ag diffraction spot intensity decays linearly with increasing the Si coverage up to 1 ML at all three temperatures. At 218 °C, this decay nearly saturates with the onset of the multilayer Si domains. Instead at 268 °C, and much more evidently at 296 °C, the Ag intensity sharply increases above 1 ML Si. This behaviour, recently observed by LEEM for the $(2\sqrt{3}\times2\sqrt{3})R30°$ phase [30], can be ascribed to a surface dewetting process, occurring when the formation of multilayer regions destabilizes the silicene layer.

The evaluation of the area covered by the different silicene phases from the diffraction spot intensities is not straightforward, as the LEED I(V) curves of the silicene structures are not identical (see Fig. 5(a)) and, particularly in the low temperature regime, the domain sizes are comparable to the transfer width of the instrument. In order to obtain a semi-quantitative estimate, we use the images in Fig. 2 to get the proportionality factors between the spot intensities of the different phases [31]. On this basis, the relative area ratios [(4×4) : ($\sqrt{13}\times\sqrt{13}$)R13.9°$a$ : ($2\sqrt{3}\times2\sqrt{3}$)R30°] at 1 ML coverage are determined to be [1.0 : 2.5 : 0] at 218 °C, [1.0 : 1.4 : 2.5] at 268 °C and [1.0 : 0 : 24.8] at 296 °C. Beyond revealing the ever-present phase coexistence, the LEED data indicate that the various silicene phases have similar structural parameters. Considering kinematic LEED, the first order diffraction spots for all phases lay at a distance r = 1.88 ± 0.04 Å$^{-1}$ from the (0 0) spot, suggesting that all structures have in-plane lattice constants closely corresponding to that of low-buckled silicene ($a_0 = 4\pi/(r\sqrt{3}) = 3.86 \approx 3.83$ Å) [1]. Accordingly, the ($\sqrt{13}\times\sqrt{13}$)R13.9°$a$ phase cannot be described as a highly compressed structure [14,19]. As seen in Fig. 5(a), the LEED I(V) curves for all phases have similar shape, with energy shift of the Bragg peaks that can be qualitatively attributed to a larger average separation between silicene and Ag with increasing the rotation angle α (see Fig. 1). This shift is especially evident for the ($\sqrt{13}\times\sqrt{13}$)R13.9°$b$ phase and may explain its instability. The nearly equal density is further supported by the growth plots of Fig. 4, in which the monolayer closure is respected by all phases simultaneously within a few percent of ML.

The above description of multi-phase silicene growth comprises several previous observations within a more extended frame. The energy and coverage dependence of the spot intensities for the different silicene phases can also explain the discrepancy in the literature on the existence of the single-phase (4×4) monolayer. Fig. 5(b-e) reports LEED patterns taken at different primary energies for 0.47 ML Si grown at 248 °C. While a multiphase pattern is clearly visible at 31 eV (Fig. 5(b)), a nearly pure (4×4) pattern is observed at 48 eV (Fig. 5(c)), in accordance with Refs. [14,19]. This energy corresponds to a Bragg minimum in Fig. 5(a), where the intensity of the ($2\sqrt{3}\times2\sqrt{3}$)R30°, ($\sqrt{13}\times\sqrt{13}$)R13.9°$a$ and $b$ phases is attenuated much more strongly than that of the (4×4) phase. Also an incomplete angular range in the LEED analysis may result in erroneous evaluations concerning the structure of the system. The coexistence of all four silicene phases, evident in Fig. 5(d), is undermined in the angular cut of Fig. 5(e). This may explain the conclusion of a uniform

(4×4) layer in Ref. [17], which bases the structural assessment on a LEED pattern similar to Fig. 5(e).

Overall, the data show that all four silicene phases have similar structures, with slight modifications due to the different adsorption configurations on the substrate. Theoretical analysis estimates the adsorption energy of silicene on Ag(111) to be in the range of a weak covalent bonding, i.e. ≈ 700 meV per Si atom. The coexistence of the silicene phases finds a basis in the density functional calculations, that predict three of the structures observed in the present study to lie within 70 meV per Si atom [23]. The calculations neglect however the effect of van der Waals interactions which, e.g. for graphene, are known to act on this energy scale [32]. For comparison, the thermal energy at the temperatures of interest is $kT ≈ 50$ meV and, therefore, sufficient to create a phase mixture during the growth process. The growth sequences show that the silicene phases follow different growth paths, which are likely to reflect both the relative cohesive energy of the phases as well as kinetic factors which control their nucleation and growth. The (4×4) structure, for instance, tends to form earlier than others, especially at high substrate temperatures, and saturates as the $(2\sqrt{3}×2\sqrt{3})R30°$ structure gains intensity at 296 °C. This behaviour is probably connected to the nucleation process of the (4×4) phase, which may initiate at the surface step edges or other defects. This assessment might indicate alternative pathways towards the creation of a uniform (4×4) silicene monolayer via a modification of the substrate morphology, for example by employing suitably stepped substrates.

Interestingly, the layer-dewetting occurring at high temperatures (268-296 °C) suggests that all silicene phases, besides being almost degenerate among themselves, are metastable with respect to the transformation into multilayer structure. The dewetting process brings into closer view kinetic factors related to the nucleation and stability of the Si structures. In this temperature range, the dewetting of the first layer closely coincides with the initial nucleation of second Si layer regions. The atomic mechanisms which trigger the dewetting need yet to be established. However, a possible scenario can envision the nucleation of small bilayer regions, such as clusters made of a very few atoms, which destabilize the surrounding single-layer neighbours and promote their aggregation to the bilayer if the thermal energy provides sufficient atomic mobility. This indicates that single-layer silicene, regardless of the phase composition, is obtained via kinetic barriers that prevent the formation of energetically more favourable multilayer regions.

By further increasing the deposition temperature, very strong morphological changes are observable also for sub-monolayer coverage [30]. Fig. 6(a) shows the LEEM image for 0.6 ML Si deposited at 315 °C. Micro-LEED analysis reveals that the light gray area is uncovered Ag (Fig. 6(b)), the dark gray area is $(2\sqrt{3}×2\sqrt{3})R30°$ silicene (Fig. 6(c)) and the white regions are multilayer

Si islands with ($\sqrt{3}\times\sqrt{3}$)R30° surface termination (Fig. 6(d)). The latter is rotated by ~ 20° with respect to the Ag(111) pattern and is, therefore, not associated with the structures described above, which have characteristic rotation angles $\alpha$ = 0°, 5.2° and 10°. LEED I(V) data for the integer ((1 0) and (0 1)) and fractional ((1/3 1/3)) spots of the multilayer island (Fig. 6(e)) are compatible with either a defective or a Ag terminated Si(111) surface (the two systems are reported to have identical LEED I(V) curves in literature) [33]. These results demonstrate that Si does not leave the Ag surface but agglomerates into three-dimensional structures with high aspect ratio, which were not observed by STM [16]. They show in this case the coexistence of multilayer islands with ($2\sqrt{3}\times 2\sqrt{3}$)R30° silicene regions, which leave largely uncovered the Ag surface. Notably, micrometer sized domains of the ($2\sqrt{3}\times 2\sqrt{3}$)R30° phase can be produced under such conditions. The permanence of only the ($2\sqrt{3}\times 2\sqrt{3}$)R30° phase at higher temperature also suggests its stronger relative stability with respect to the other silicene structures.

Lastly, we notice that our results have strong impact in the interpretation of several microscopic and macroscopic properties of silicene. For instance, the electronic band structure, the chemical reactivity, the atomic mobility, and the thermodynamics of the system depend on the distribution, size, and structure of the different phase domains. Similarly to graphene, the presence of rotational domains and related boundaries, could sizably reduce the in-plane electron mobility of silicene, if grown on non metallic substrates, and alter the transport behaviour predicted for uniform silicene layers [34]. Indeed, the coexistence of different silicene phases should be apparent when macroscopic probes are used to investigate the electronic structure of the Si/Ag(111) system.

## 4. Conclusions

In conclusion, by means of LEED and LEEM studies, we determined the conditions that allow fine control of the silicene growth on Ag(111), with defined relative phase abundances. While micrometer sized silicene domains are identified, complete single-phase monolayers are not observed under any of the examined conditions. The results highlight a rich phase formation diagram, comprising distinct, but almost degenerate, silicene phases. Synthesis of these phases can be achieved under kinetically limited conditions, while competing processes lead to Si multilayer formation and surface dewetting.

**Figure captions**

Figure 1: Structural models and corresponding LEED patterns for different silicene phases. White and orange circles represent Si and Ag atoms. The green rhombi define the unit cells of the silicene superstructures. Red and black circles indicate the silicene and Ag(111) diffraction spots. (a,e) (4×4), α = 0°. (b,f) ($\sqrt{13}\times\sqrt{13}$)R13.9°*a*, α = 5.2°. (c,g) ($2\sqrt{3}\times2\sqrt{3}$)R30°, α = 10°. (d,h) ($\sqrt{13}\times\sqrt{13}$)R13.9°*b*, α = 27°. Arrows indicate the first order diffraction spots of the silicene phases.

Figure 2: 1.3 ML Si grown on Ag(111) at 248 °C. (a) LEED pattern at 31 eV. Spots of the ($\sqrt{3}\times\sqrt{3}$)R30° pattern are indicated by a white arrow. (b-d) Dark field LEEM images obtained by selecting the first order diffraction spots corresponding to (b) the (4×4) structure and (c,d) the two rotational domains of the ($\sqrt{13}\times\sqrt{13}$)R13.9°*a* structure. All three dark field images are acquired at 31 eV. (e) Sum of the three dark field images after drift correction. (f) Bright field LEEM image at 14 eV.

Figure 3: LEED pattern of Si/Ag(111) at monolayer coverage after growth at (a) 163 °C, b) 218° C, (c) 232 °C, (d) 248 °C, (e) 268 °C, (f) 296 °C. The outermost intense spots belong to the substrate. Electron energy is 31 eV.

Figure 4: Coverage-temperature phase diagram illustrated by LEED snapshots during growth at different temperatures. Each row is a growth sequence at the temperature noted on the left. The evolution of the relevant spot intensities is given in the plots on the right for each temperature. All data are acquired at an electron energy of 31 eV.

Figure 5: (a) LEED I(V) plots from the primary LEED spots of each silicene phase on Ag(111). The curves are normalized to the peak intensity of the pronounced Bragg reflection around 30 eV. All patterns appear 6-fold symmetric, and the intensities reflect the average over six spots. The energy-dependence of the relative spot intensities is illustrated for 0.47 ML Si grown at 248 °C at electron energies of (b) 31 eV, (c) 48 eV and (d,e) 27 eV.

Figure 6: (a) LEEM image after the growth of 0.6 ML Si at 315 °C. Electron energy is 8 eV. Microspot-LEED data demonstrate that (b) light gray areas correspond to uncovered Ag(111), (c) dark gray areas to ($2\sqrt{3}\times2\sqrt{3}$)R30° silicene, and (d) few bright regions to multilayer Si(111)

crystals. (e) Microspot-LEED I(V) data from a thick Si crystal show the energy dependence of the principal spots for the ($\sqrt{3}\times\sqrt{3}$)R30° reconstructed Si(111) surface.

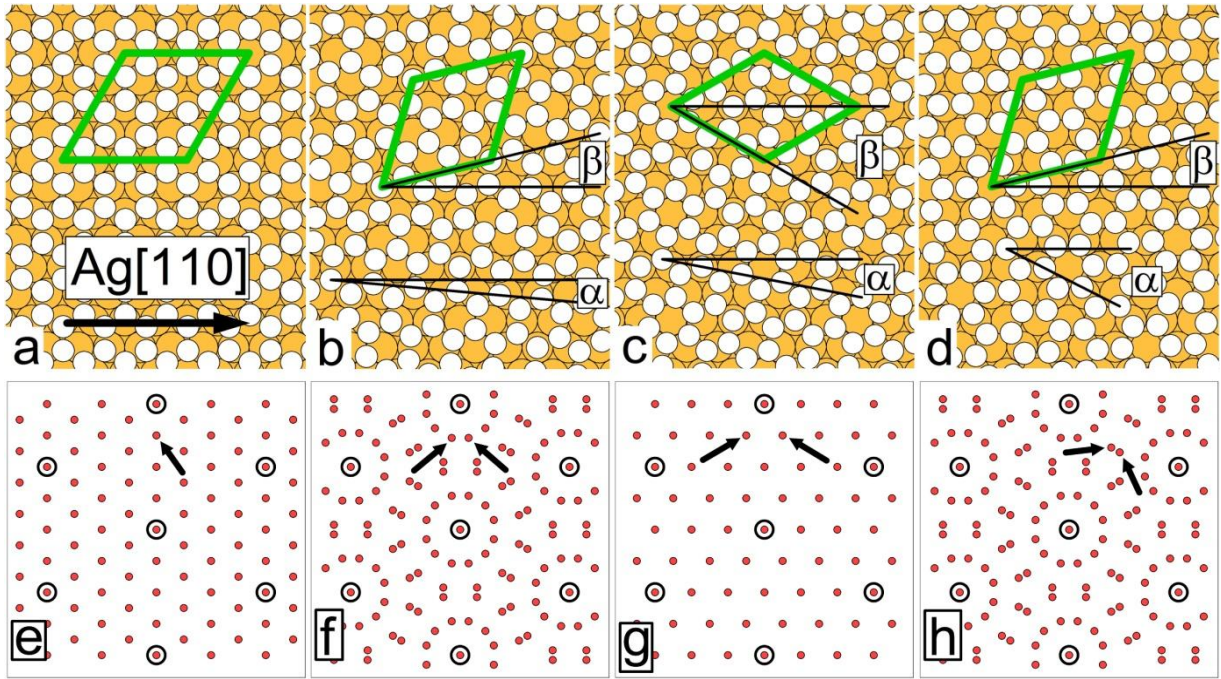

Figure 1

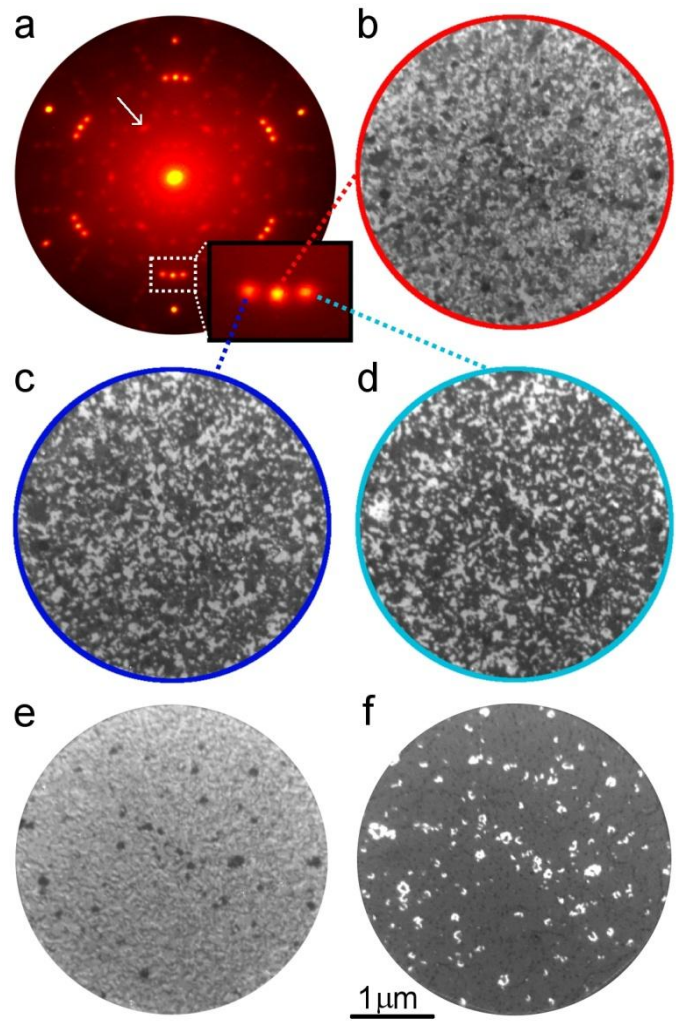

Figure 2

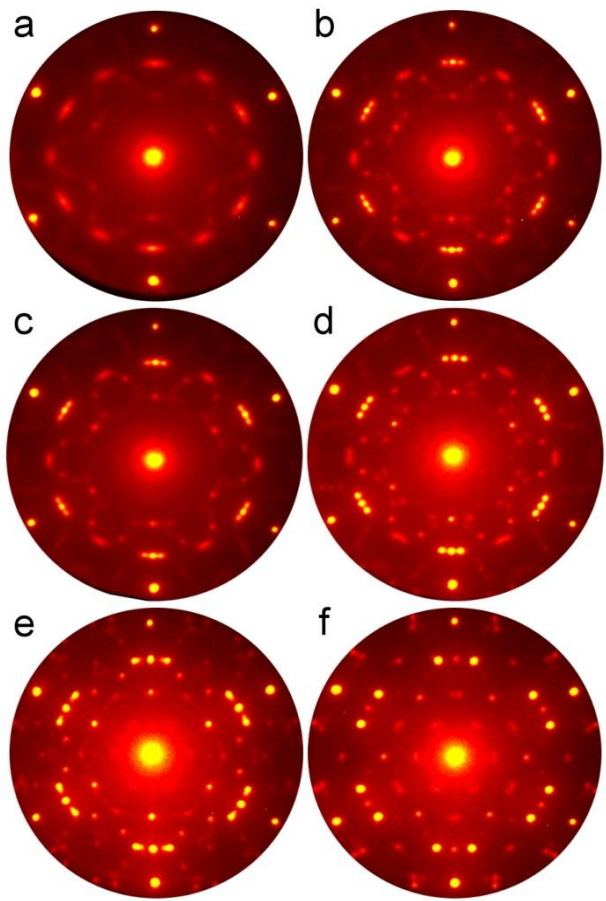

Figure 3

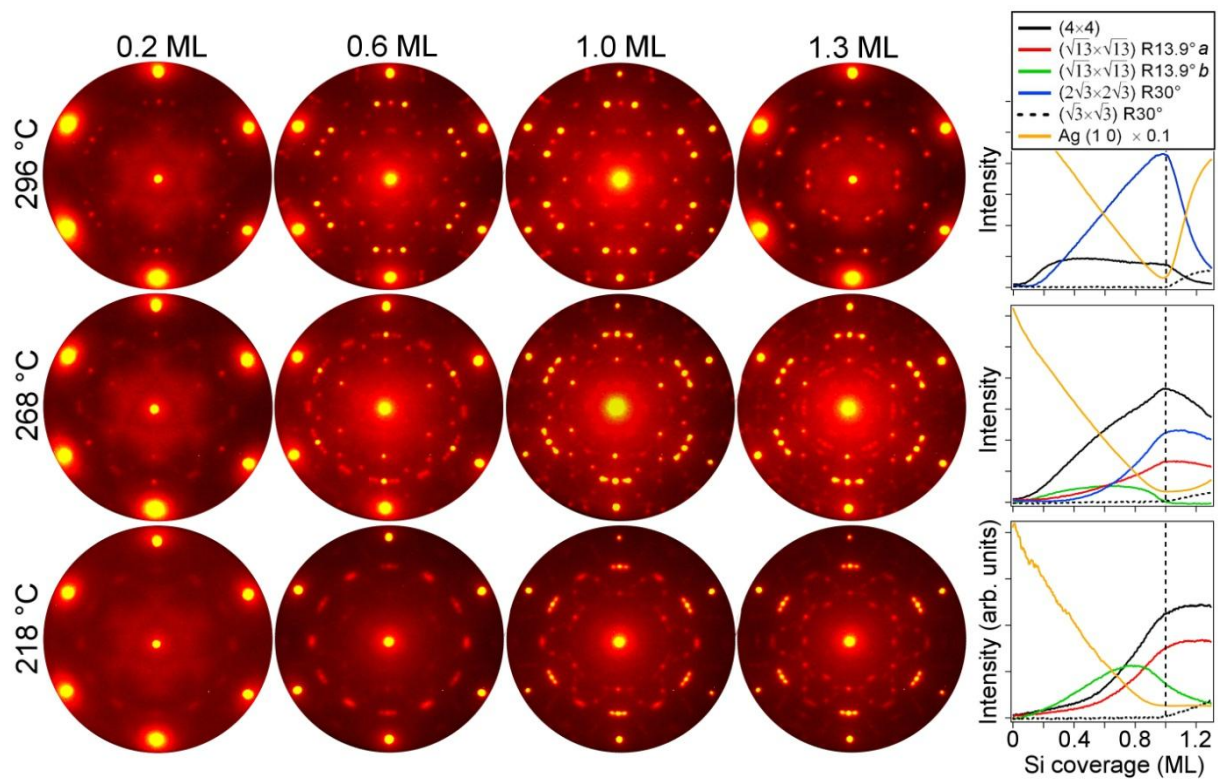

Figure 4

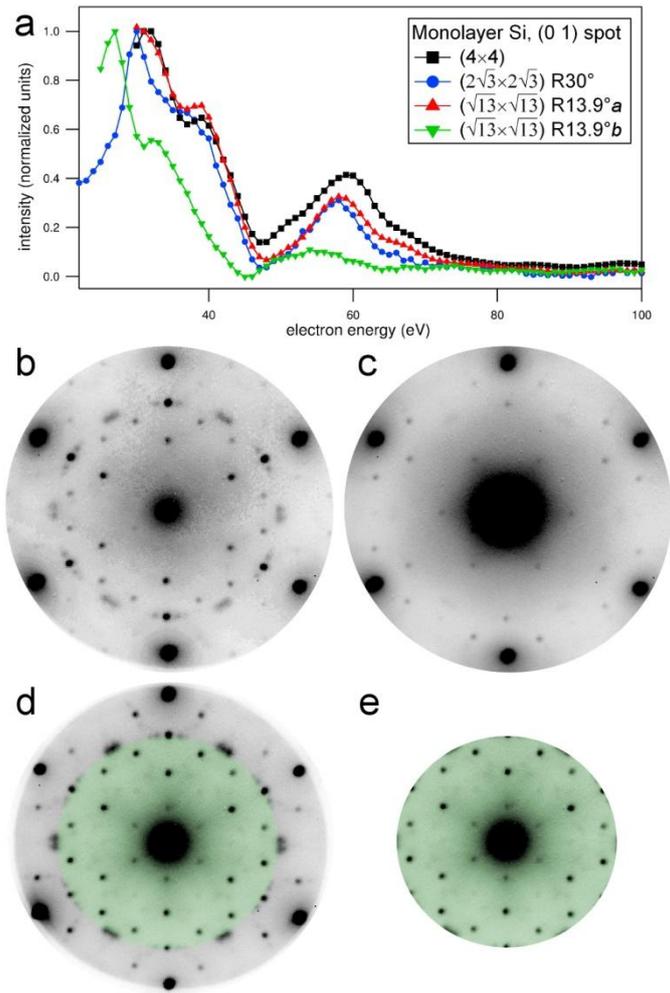

Figure 5

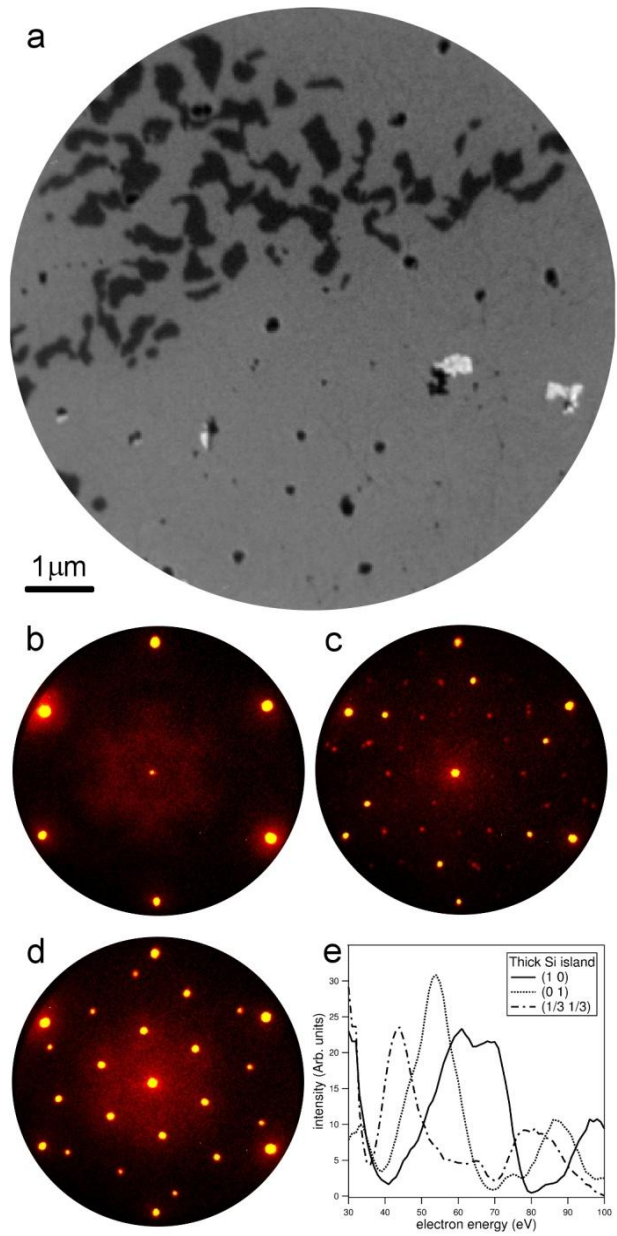

Figure 6